\documentclass{article}

\usepackage{arxiv}

\usepackage[utf8]{inputenc} % allow utf-8 input
\usepackage[T1]{fontenc}    % use 8-bit T1 fonts
\usepackage{hyperref}       % hyperlinks
\usepackage{url}            % simple URL typesetting
\usepackage{booktabs}       % professional-quality tables
\usepackage{amsfonts}       % blackboard math symbols
\usepackage{amsmath}
\usepackage{nicefrac}       % compact symbols for 1/2, etc.
\usepackage{microtype}      % microtypography
\usepackage{lipsum}
\usepackage{graphicx}
\graphicspath{ {./images/} }

\usepackage[
backend=biber,
style=numeric,
sorting=ynt
]{biblatex}
\title{Chaotic oscillator networks for classification tasks.}

\author{
 Ivas T\\
  Intelligent Manufacturing Group (IMG) \\ 
  Empa, Thun, Switzerland \\
  \texttt{toni.ivas@empa.ch} \\
   \And
  Violakis G. \\
Hellenic Mediterranean University \\ 
Heraklion, Greece \\
\texttt{violakisg@hmu.gr}
  \And
Richter R.\\
Intelligent Manufacturing Group (IMG) \\ 
  Empa, Thun, Switzerland \\
  \texttt{roland.richter@empa.ch} \\
  \And
Hoffmann P. \\
Intelligent Manufacturing Group (IMG) \\ 
  Empa, Thun, Switzerland \\
  \texttt{patrik.hoffmann@empa.ch} \\
 \And
Shevchik S.A. \\
Intelligent Manufacturing Group (IMG) \\ 
  Empa, Thun, Switzerland \\
  \texttt{sergey.shevchik@empa.ch} \\
}

\begin{document}
\maketitle
\begin{abstract}
Chaotic oscillators have gained significant attention in the research community because of their ability to reproduce and investigate the complex dynamics of real-world phenomena. Recent advances in the design of chaotic oscillator ensembles have led to the development of efficient signal processing frameworks that surpass traditional approaches. However, scaling such systems remains challenging due to the significant increase of computational resources and issues with training convergence. This study advances the state of the art by addressing the problem of data processing with ensembles of nonlinear oscillators that can be scaled up. In our approach, the processing is achieved as an anticipated local resonance or echo in a group of coupled chaotic oscillators, driven by external data input. Local resonance is enabled by tuning the coupling terms between the oscillators, which are approximated using the traditional artificial neural network and adapted to match the input feature distributions. Training the framework entails  training this  neural network to capture the dynamics of the entire oscillator system. The framework is evaluated using synthetic data and demonstrates an accuracy in machine learning classification task, while  patterns recognition and dynamic system identification are also presented here as an extension of the functionality that involves additional modifications. Additionally, the universality of this approach is demonstrated by tests with different connections configurations between the oscillators and their types. The main advantage  of the proposed framework is that it avoids hand-crafting explicit coupling terms, which requires expert knowledge and does not scale for large problems. Leveraging standard machine learning components simplifies both training and deployment of oscillator networks, enabling gradient-based optimization. 
\end{abstract}

% keywords can be removed
%

\section{Introduction}
Chaotic oscillations appear in a wide variety of biological, physical, and engineering systems \cite{kapitaniak_chaotic_1992}. For a long time, the phenomenon was poorly understood \cite{kapitaniak_chaotic_1992} until Lorenz's groundbreaking work \cite{norton_deterministic_1963} showed how its dynamics could be explicitly recovered. Since then, chaos has attracted significant attention from the research community, spurring the rapid development of numerous theoretical and applied concepts \cite{kapitaniak_chaotic_1992,skiadas_applications_2016}. In recent years, chaos-based approaches have carved out a niche in practical applications, offering advantages over traditional methods, particularly when dealing with extremely low signal-to-noise ratios and unraveling complex data behaviors \cite{kapitaniak_chaotic_1992,skiadas_applications_2016}. 
However, practical examples of classical chaos-based algorithms remain scarce, with applications mainly limited  to areas such as data encryption \cite{pushpalatha_collective_2024}, data transfer \cite{kaddoum_wireless_2016,alkhonaini_two-phase_2024}
, and routing \cite{luo_improved_2024}. The success of these applications, combined with the ability to build complex functionality, scale it up, and track data dynamics,  motivates us to consider chaos as a viable framework for next-generation machine learning (ML). A variety of chaos-based ML approaches have been proposed, including mimicking the human brain via neuronal spiking \cite{nobukawa_synchronization_2020}, introducing global self-organization within artificial neural networks \cite{landmann_self-organized_2021}, and improving training procedures. For detailed reviews with additional examples, we refer the reader to \cite{harikrishnan_novel_2019,akgul_chaos_2024}. Despite the enhanced functionality that chaotic systems can offer, their broader adoption in practice remains limited. A key reason is the substantial computational cost required to support sufficiently large models, ensure training convergence, and accuracy-- particularly on larger datasets. The core challenges in chaotic systems include stabilization and synchronization issues \cite{kapitaniak_chaotic_1992}. In this paper, we highlight these challenges for the networks under investigation, while a more general treatment of chaos dynamics can be found in the literature \cite{kapitaniak_chaotic_1992,lynch_chaos_2018}. 

The first formulations of chaos synchronization and stabilization emerged in control theory \cite{ott_controlling_1990} and were later extended to distributed systems \cite{ding_recovering_nodate}, including chaotic networks. Distributed systems add complexity by requiring global dynamics to emerge from multiple locally controlled -- or uncontrolled -- chaotic states,  which govern interaction among neighboring nodes across different spatial and temporal scales \cite{ding_recovering_nodate, ling_delay-dependent_2021}. The coupling terms that describe these interactions are often nonlinear, further complicating the analysis and design of such systems \cite{ding_recovering_nodate}. 
Despite these challenges, several efficient classical chaos control methods have been proposed. Ling et.al \cite{ling_delay-dependent_2021} developed delay pinning impulsive control to suppress the chaos in large-scale networks. Zhao et.al. \cite{zhao_finite-time_2021} introduced a finite–time control scheme for networks governed by differential-differences equations. Gharb et.al \cite{ghrab_new_2021} applied time sliding mode control to distributed networks, while Luo et.al. \cite{luo_event-triggered_2021} presented finite-time synchronization control for networks with reaction-diffusion coupling. Serrano et.al. \cite{serrano_robust_2022} proposed an asynchronous control strategy effective delay dependent networks.

 Despite their demonstrated efficacy, these methods typically require detailed a priori knowledge of underlying network dynamics (or at least their qualitative form), which is not always available. Moreover, even minor changes to node properties, connectivity patterns, or coupling functions often necessitate a complete redesign of stabilization and synchronization parameters.
In this context, ML offers a powerful alternative for chaos stabilization and synchronization \cite{garcia_machine_2022,huawei_long-term_2020}. As a universal approximator, ML can infer  chaotic dynamics,directly from observed input output data, eliminating the need for explicit knowledge of underlying network dynamics. In addition, the scalability of deep neural networks makes it possible to  and analyze highly complex chaotic behaviors \cite{garcia_machine_2022,huawei_long-term_2020}, and even to drive their synchronization \cite{mobini_deep_2020} and stabilization \cite{cheng_deep_2023}.
This study uses machine learning (ML) to stabilize and synchronize chaotic networks designed for data processing tasks, with classical oscillators serving as network nodes. We investigate several well-known chaotic oscillators as alternatives to traditional artificial neurons. These oscillators offer multiple advantages, including of straightforward implementation in low-cost electronic circuits \cite{chua_chaos_1993}, optical elements \cite{pai_experimentally_2023} or memresistor networks \cite{ye_overview_2022}. These networks can be efficiently modeled in FPGA/CPU/GPU architectures \cite{guillen-fernandez_synchronization_2019}. They also offer relatively low computational costs and the ability to realize the full spectrum of chaos-based phenomena across different spatial-temporal scales \cite{kapitaniak_chaotic_1992}.
Beyond synchronization and stabilization, data processing in these systems relies on local chaotic resonance. In this phenomenon, oscillation modes across the network's spatial domain can concentrated locally for brief periods, amplifying oscillations in individual nodes. This effect is well described in classical \cite{chirikov_universal_1979} and more recent studies \cite{d_multiple_2024}, and in our approach it is induced by designing the coupling terms of oscillators in response to the input pattern. It is important to note that our study builds on the ideas of Izhikievich and colleagues  \cite{hoppensteadt_synchronization_2000, hoppensteadt_pattern_2000,izhikevich_weakly_1999} and extends them further. We also acknowledge two closely related lines of work that motivated this research. 
Huang et.al. \cite{huang_self-organizing_2015} introduced self-organizing map (SOMs) based on short-term cycle attractors with artificial neurons as network nodes. Their findings regarding the enhanced processing capabilities of SOMs motivated us to extend our work by utilizing oscillators as network nodes. Similarly, the work of Wang et.al. \cite{wang_training_2024} proposed a specialized form of equilibrium propagation to train networks of phase oscillators for pattern recognition. In their work, the coupling terms of Kuramoto oscillators are learned with machine learning, yielding distinct network response to the learned patterns. While our study partially parallels their study, we employ standard backpropagation training algorithms for the machine learning component and evaluate applicability across multiple topologies and oscillator types.
This paper is organized as follows. Methods describes the framework architecture and the governing equations for each processing step. Results and discussion presents the performance of on classification and system identification tasks. Conclusion summarizes our contributions and outlines future directions.

% Methods section
\section{Methods}
\label{sec:headings}
This work investigates networks of chaotic oscillators for classification and pattern recognition, using machine learning to approximate the coupling terms between nodes so that the network exhibits the desired local chaotic resonance in response to learned data patterns.We explore several network designs that employ classical oscillators as nodes for pattern recognition tasks. The objective is to induce local chaotic resonance for specific input patterns by learning the nodes coupling terms with ML. The investigation focuses on two specific classical oscillators: {\em FitzHugh-Nagumo}, and {\em Kuramoto}, which are characterized by different parameter sets and coupling, with the parameter list shown in Table \ref{tab:fhn} below (see Additional information for full list).

\begin{table}[h]
    \centering
    \begin{tabular}{ccc} \hline
        $\epsilon$        & $0.05$                    & Time-scale separation (fast/slow)            \\ \hline
        $a$               & $0.5$                     & Threshold / excitation parameter            \\ \hline
        $\sigma$          & $0.5,\;0.06,\;0.72$       & Diffusive coupling / noise strength         \\  \hline
        $\sigma_s$        & $0.006,\;0.1$             & Spike-rate scaling constant                  \\  \hline
        $T_{\text{span}}$ & $(0,\;300)$               & ODE integration time window                  \\  \hline
         &  & \\
    \end{tabular}
    \caption{Parameters used in the simulation of the FitzHugh-Nagumo oscillator networks.}
    \label{tab:fhn}
\end{table}
\paragraph{Classical chaotic oscillators as network nodes.}
{\em FitzHugh-Nagumo} (FHN) oscillator network is a prototype model of neural dynamics \cite{fitzhugh_impulses_1961}. Unlike the classical Duffing oscillator, where the energy is injected by an external force $F$, the {\em FitzHugh-Nagumo} system stores energy internally and releases it in response to an external stimulus, after which it relaxes along a limit cycle to a minimum energy state. In the literature, this model is often described as an excitable membrane \cite{al_beattie_criticality_2024}, and its general dimensionless form is given by:
\begin{align}
 \label{eq:fhn}
    \epsilon_0 \frac{du_k}{dt} &= j_0 + u_k - \frac{u_k^3}{3}-v_k + \sigma \sum_{j=1}^{N} W_{kj}
    [B_{uu}(u_j-u_k) + B_{uv}(v_j-v_k)] \\
\label{eq:rec_var}
    \frac{dv_k}{dt} &= e_0 + u_k + a \sigma \sum_{j=1}^{N} W_{kj}[B_{vu}(u_j-u_k) + B_{vv}(v_j-v_k)]
\end{align}

Here, $k = 1\text{..}N$, $a$ is the exciting threshold; $u$ denotes the membrane potential, $v$ is a recovery variable, and $\epsilon_0$ sets the time scale separation between excitation and relaxation cycles. The overall coupling between nodes is scaled by the coupling strength $\sigma$. The interaction between membrane potential and the recovery variable is governed by a $2 \times 2$ rotation matrix $B$, which is defined by a single angle $\phi$:
\begin{equation}
\label{eq:B_matrix}
B = \begin{pmatrix}
    \cos(\phi) & \sin(\phi) \\
    -\sin(\phi) & \cos(\phi)
\end{pmatrix}
\end{equation}
As shown above, Eq.(\ref{eq:fhn}) governs the evolution of the membrane potential, while Eq.(\ref{eq:rec_var}) describes the recovery dynamics following each excitation-relaxation cycle. The system's excitation versus idle states are determined by the parameter $a$, where $|a|> 1$ corresponds to the excitable state, whereas $|a|< 1$ yields the self-sustained limit cycle (idle) state \cite{majhi_chimera_2019}. The dynamics of {\em FitzHugh-Nagumo} oscillator, controlled by the parameters $a$, $\phi$, $v$, can range from periodic to chaotic behavior. A comprehensive overview of the possible regimes is provided in \cite{cebrian-lacasa_six_2024}, underscoring the importance of parameter selection for network dynamics.  While in present study those oscillators were kept in chaotic mode or at edge of chaos \cite{cebrian-lacasa_six_2024} (see exact parameters in Table I). Data input was implemented by enforcing changes in the membrane potential in Eq.(\ref{eq:fhn}) proportional to the corresponding values of the input data.
We choose {\em FitzHugh-Nagumo} oscillator due to its widespread use in the life sciences, including spiking neurons models \cite{benson_multi-scale_2021}, reservoir computing \cite{al_beattie_criticality_2024}, and cardiac tissue potential propagation \cite{benson_multi-scale_2021}. By coupling {\em FitzHugh-Nagumo} oscillators under various topologies, we draw analogy to brain circuits, highlighting the potential of our methods for modeling biological systems in general and neuronal circuits in particular. 
{\em Kuramoto} model defines a system of $N$-coupled oscillators, where the state of each oscillator is defined as \cite{odor_critical_2019} :
\begin{equation}
\label{eq:kuramoto}
\frac{d\phi_i}{dt} = \omega_i + \frac{1}{N}\sum_{j=1}^{N} k_{ij}\sin(\phi_i - \phi_j) + h \sin(\phi_i-\psi_i)
\end{equation}
In Eq.(\ref{eq:kuramoto}), the oscillation is driven by the intrinsic eigenfrequency $\omega_k$,  influenced by neighboring oscillators that induce phase shifts $\phi_i$, where $k_{ij}$ represents the connection weight between oscillators $i$ and $j$. The chaotic dynamics in Eq.(\ref{eq:kuramoto}) emerges from the complex interplay of these phase shifts \cite{odor_critical_2019}, which can lead to phenomena, such as phases self-synchronization and firing regimes similar to those observed in excitable systems \cite{odor_critical_2019}. Like the {\em FitzHugh-Nagumo}, {\em Kuramoto} model  is widely used to simulate neuronal responses to external stimuli \cite{odor_critical_2019}.

\begin{figure}
  \centering
   \includegraphics[width=0.8\linewidth]{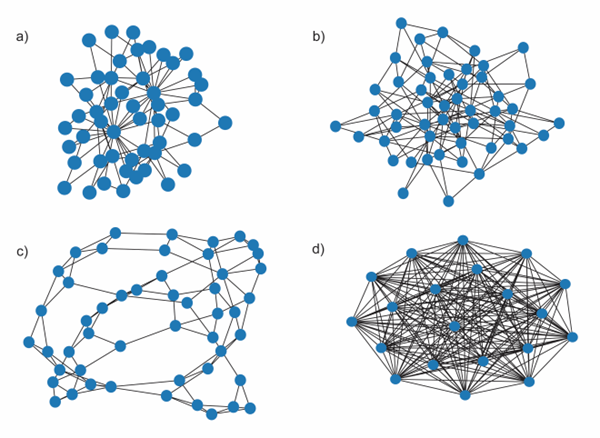}
  \caption{Tested network topologies with a) Albert-Barnabasi graph, b) Erd\H{o}s--R\'enyi (random) graph, c) Watts-Strogatz (small-world graph) , d) full connectivity used for classification tests.}
  \label{fig:fig1}
\end{figure}

\begin{figure}
  \centering
    \includegraphics[width=0.8\linewidth]{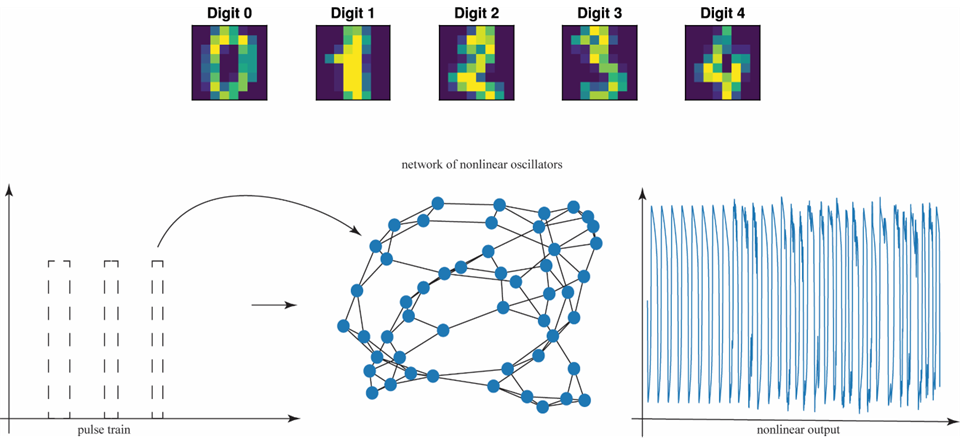}
    \caption{Network feeding and training schemes: a) The examples from scikit-learn dataset pixel intensity is converted to the pulse-train b) The pulse-train is fed to the network of nonlinear oscillators with specific topology and nonlinear output is trained using the 3-layer neural network. The pattern in (b) corresponds to scikit-learn digits patterns. The coupling in (b) corresponds to $\Phi$ in Eq.(5); c) network training with reservoirs.}
  \label{fig:fig2}
\end{figure}

\paragraph{Oscillators coupling and networks topology}
Like Eq.(\ref{eq:kuramoto}), the system of coupled oscillators in Eqs.(\ref{eq:fhn}) - (\ref{eq:rec_var}), and (\ref{eq:kuramoto}) can be described by an additional coupling term, which characterizes the mutual influence of the oscillators. With this extension, the general case of $N$-coupled oscillators is defined as:  
% equation for coupled oscillators
\begin{equation} \label{eq:coup_oscil}
   \begin{matrix}
    \ddot{x}_1 =  \alpha \dot{x}_1 + \beta x_1 +\gamma x_1 + \Phi_1(x_1,x_2,...,x_N) \\
    ...\\
    \ddot{x}_N =  \alpha \dot{x}_N + \beta x_N +\gamma x_N + \Phi_N(x_1,x_2,...,x_N) \\
    \end{matrix} \Longrightarrow
    \begin{matrix}
    \ddot{x}_1 =  \alpha \dot{x}_1 + \beta x_1 +\gamma x_1 + NN(x_1,x_2,...,x_N;\theta) \\
    ...\\
    \ddot{x}_N =  \alpha \dot{x}_N + \beta x_N +\gamma x_N + NN(x_1,x_2,...,x_N;\theta) \\
    \end{matrix}
\end{equation}
Here, $NN(x_1,x_2,...,x_N; \theta)$ is the neural network with trainable parameters $\theta$, which represents the coupling term $\Phi_i$ a nonlinear function that explicitly describes the mutual interactions between oscillators.  For the FitzHugh-Nagumo model, the coupling term $\Phi_i$ is applied only to Eq.(\ref{eq:fhn}), since Eq.(\ref{eq:rec_var}) governs the intrinsic properties of the oscillator, specifically, its recovery, and was therefore kept constant for the entire experiments duration.  
The coupling term $\Phi$ in Eq.(\ref{eq:coup_oscil}) defines the network topology, taking a zero value for unconnected oscillators and acting as an interaction function for connected ones. In our study, the case of directional independence of interaction propagation across the network was considered. Additionally, for some considered network topologies holds: $\Phi_1=\Phi_2=\text{...}=\Phi_N$, with the corresponding example in Figure \ref{fig:fig1}, d. The results presented here are limited to a selection of network topologies, shown in Figure \ref{fig:fig1}, a)-d). We tested several topologies using our framework, provided as the code repository. Not all results are reported here due to space limits, however the cases from Figure \ref{fig:fig1}, a)-d) showed similar dynamic properties. The biological networks usually are modeled using small-world network models \cite{watts_collective_1998}. In this work we use a Watt-Strogatz model, that is small-world network with mean degree of $2k$, using $n$ nodes and $nk$ edges which is shown in Fig. \ref{fig:fig1} c). Other network topologies used are full connectivity that showed good characteristics in training of the XOR gate as reported by
With all mentioned above, the architecture presented in this study replaces the explicit coupling $\Phi$ (see Eq.(\ref{eq:coup_oscil}), left term) by machine learning approximator $NN(x_1,x_2,...,x_N;\theta)$(see Eq.(\ref{eq:coup_oscil}), right term). The tests of this approach are verified on various network topologies and types of oscillators. Additionally, simple classical ML training techniques are employed here for network tuning. 
\paragraph{Network data feeding} \label{sec:network_data_feed}
In our study, we present  two classification cases using two nonlinear oscillator networks specifically {\em FitzHugh-Nagumo} (FHN) and {\em Kuramoto networks (KN)}. In both cases, we employ nonlinear oscillators in reservoir computing framework and train only the output layer for the classification tasks. Following Al Beattie et. al. \cite{al_beattie_criticality_2024}, we encode the each pixel's intensity as short pulse train and feed this into the nonlinear network. 
The first dataset is the scikit-learn digits image patterns, as illustrated in Figure \ref{fig:fig2} a). In our setup, the number of oscillators is chosen to satisfy condition given by reservoir computing literature \cite{al_beattie_criticality_2024,bollt_explaining_2021}. We feed the data by converting each pixels intensity into proportional, short lived perturbation of its corresponding oscillator, applied independently of oscillator's current state. A schematic illustration of pixel-wise input scheme into the network is shown in Figure \ref{fig:fig2} b). The inputs are delivered as spike trains that drive the network via external forcing, after which the system is allowed to evolve until it reaches a equilibrium. The network's dynamics during this free evolution are interpreted as the classification response. 
The input procedure for each oscillator in the network is as follows:
For the FitzHugh-Nagumo oscillator, the perturbation is a sudden change in membrane potential $\mu$, as defined in Eqs.(\ref{eq:fhn}) and (\ref{eq:rec_var}). Note that the external stimulus is held constant, while $u$ is an intrinsic property of the oscillator that shapes its free motion dynamics.
In the case of the {\em Kuramoto} model, the perturbation is applied by adjusting the phase $\phi_i$, as specified in Eq.(\ref{eq:kuramoto}).

\paragraph{Network training} \label{sec:Network_training}
We employed two different training strategies.
Prior training, the expected response of the oscillators network to the given image patterns is generated and presented (see the expected dynamics in Figure \ref{fig:fig2}, b). This procedure involves assigning different combinations of resonant and non-resonant oscillators, whose collective dynamics yield a unique response to each input pattern (i.e. effectively performing a single classification act). Subsequently, we trained an artificial neural network to approximate the coupling terms in the oscillator equations as defined in Eq.(\ref{eq:coup_oscil}) with the objective of reproducing the prescribed target response for given inputs. This formulation closely follows the Universal Differential Equations (UDEs) paradigm of Rackauckas et. al \cite{rackauckas_universal_2021}, wherein unknown components of the governing dynamics in our case the coupling terms are represented by a neural network and learned form data.

%% training ESN or RC 
\paragraph{Echo State Networks}
 A Echo State Network (ESNs) represent a simplification of a recurrent neural networks, as introduced by Jaeger \cite{noauthor_jaeger_nodate, jaeger_harnessing_2004}. In ESNs or RC computing the training includes reservoir of the nonlinear oscillator nodes which are randomly connected following one of the network topologies illustrated in Figure \ref{fig:fig1}. Data are sequentially fed to the external inputs of the reservoir nodes. The actual dynamics of the oscillator network are then compared with the expected dynamics, enabling optimization of the output neural network to achieve desired solution.
 This training is organized using reservoirs, as illustrated in Figure \ref{fig:fig2}, c). Our preference for reservoir training over standard methods is motivated primarily by the use of {\em FitzHugh-Nagumo} oscillators. In particular,  recent advancements in reservoir computing demonstrate that {\em FitzHugh-Nagumo} oscillators are robust to parameter variations and achieve high accuracy in classification tasks. Numerous studies in the literature have explored the computational capabilities of reservoirs operating near the chaotic regime. The primary goal of a reservoir is to perform a deterministic mapping between input and output. In recent years, there has been growing interest in applying reservoir computing to predict dynamical systems. For a given training dataset, the RC can be defined as:
\begin{align}
    r_{i+1} &= (1-\alpha)r_i + \alpha q(A r_i + u_i + b) \\
    y_{i+1} &= W^{out} r_{i+1}
\end{align}
where reservoir variable $r_i \in R^{d_r}$ is chosen to have a much higher dimensionality then the input data, such that $d_r\>>d_x$. Here, $q$ is nonlinear function; in our case, we select $swish$ function. The $\alpha$ parameter gives mixing between previous and current states of RC system, though it is often set to $1$. The lifting operation is defined as in \cite{bollt_explaining_2021}: 
\begin{equation}
    u_i = W^{in}x_i
\end{equation}
% figure
Where $W^{in}$ is randomly initialized matrix with dimensions ($d_r,d_x$). The matrix $A$ is a square matrix designed to exhibit specific properties, including a spectral radius close to $1$ and sparse structure \cite{lu_attractor_2018, pathak_model-free_2018}. The readout matrix $W^{out}$ provides a linear transformation, that is trained to map the input data $x_i$ to the target data $y_{i+1}$. This represents a major advantage of reservoir computing, as $W^{out}$ can be obtained through simple ridge least-square optimization \cite{bollt_explaining_2021}. 

\paragraph{Equilibrium Propagation} %%checked
Equilibrium Propagation (EP) is a learning framework introduced by Scellier and Bengio \cite{scellier_equilibrium_2017} for energy-based models. Unlike the standard backpropagation algorithm, which requires a special computational circuit during the second phase of training the computational graph, EP offers the advantage of using a single computational circuit for both inference and training. Given that nonlinear oscillator networks have well defined
energy functions and exhibit relaxation-type dynamics, EP can be effectively applied to train these networks.
Following Eq.(\ref{eq:kuramoto}), the total energy of the system can be expressed as:
\begin{equation}
\frac{d\phi_i}{dt} = -\frac{\partial F}{\partial \phi_i}(\phi,\theta,\beta)
\end{equation}

where the free energy $F(\phi,\theta,\beta)$ is defined as:
\begin{equation}
    \label{eq:free_ene_kuramoto}
    F = E(\phi;\theta) + \beta C(\phi_{out},\phi^{\tau})
\end{equation}
where $\theta$ are trainable parameters, and $C(\phi_{out}, \phi_{\tau})$ is cost function defined by difference between output variable $\phi_{out}$ and target $\phi_{\tau}$.
The gradient of the cost function with respect to trainable parameters $\theta$ can be calculated as:
% --- EP gradient approximation ------------------------------------
\begin{equation}
    \frac{\partial C}{\partial\theta_\alpha}
     \;=\;
     \left.\frac{d}{d\beta}
            \!\left(
                \frac{\partial F}{\partial\theta_\alpha}
            \right)\right|_{\beta = 0}
     \;\;\approx\;\;
     \frac{1}{\beta}\Bigl(
          \bigl\langle\frac{\partial E}{\partial\theta_\alpha}\bigr\rangle_{\text{nudge},\,\phi^{\text{in}}}
          \;-\;
          \bigl\langle\frac{\partial E}{\partial\theta_\alpha}\bigr\rangle_{\text{free},\,\phi^{\text{in}}}
     \Bigr).
\end{equation}
According to Eq.(\ref{eq:kuramoto}) the analytical derivatives of energy function are given by:
% --- Analytical derivatives of the energy -------------------------
\begin{align}
\frac{\partial E}{\partial k_{ij}}&=
    -\cos\bigl(\phi_i - \phi_j\bigr) \\
\frac{\partial E}{\partial h_i}&=
    -\cos\bigl(\phi_i - \psi_i\bigr) \\
\frac{\partial E}{\partial \psi_i}&=
    -h_i\sin\bigl(\phi_i - \psi_i\bigr).
\end{align}
Recent work by Wang et al. \cite{wang_training_2024} has established coupled phase oscillators (such as {\em Kuramoto model}) as a viable neuromorphic
platform for supervised learning.
\paragraph{Coupling network} %%checked
The connections in the network of nonlinear oscillators are modeled based on biological analogies. For instance, axonal connections in mammals are unidirectional and involve a small delay due to chemical interactions. However, some organisms, such as Hydra exhibit bidirectional axonal connections; in our study, we focused solely on unidirectional connections. The mean value of the these connections serves as a control parameter that determines the criticality within the network. We modeled the connection weights in our networks using a standard Gaussian distribution. As in previous studies, we did not incorporate delays into the networks, unlike what occurs in the biological systems. 

\section{Results and discussions}
\begin{figure}
    \centering
    \includegraphics[width=0.8\linewidth]{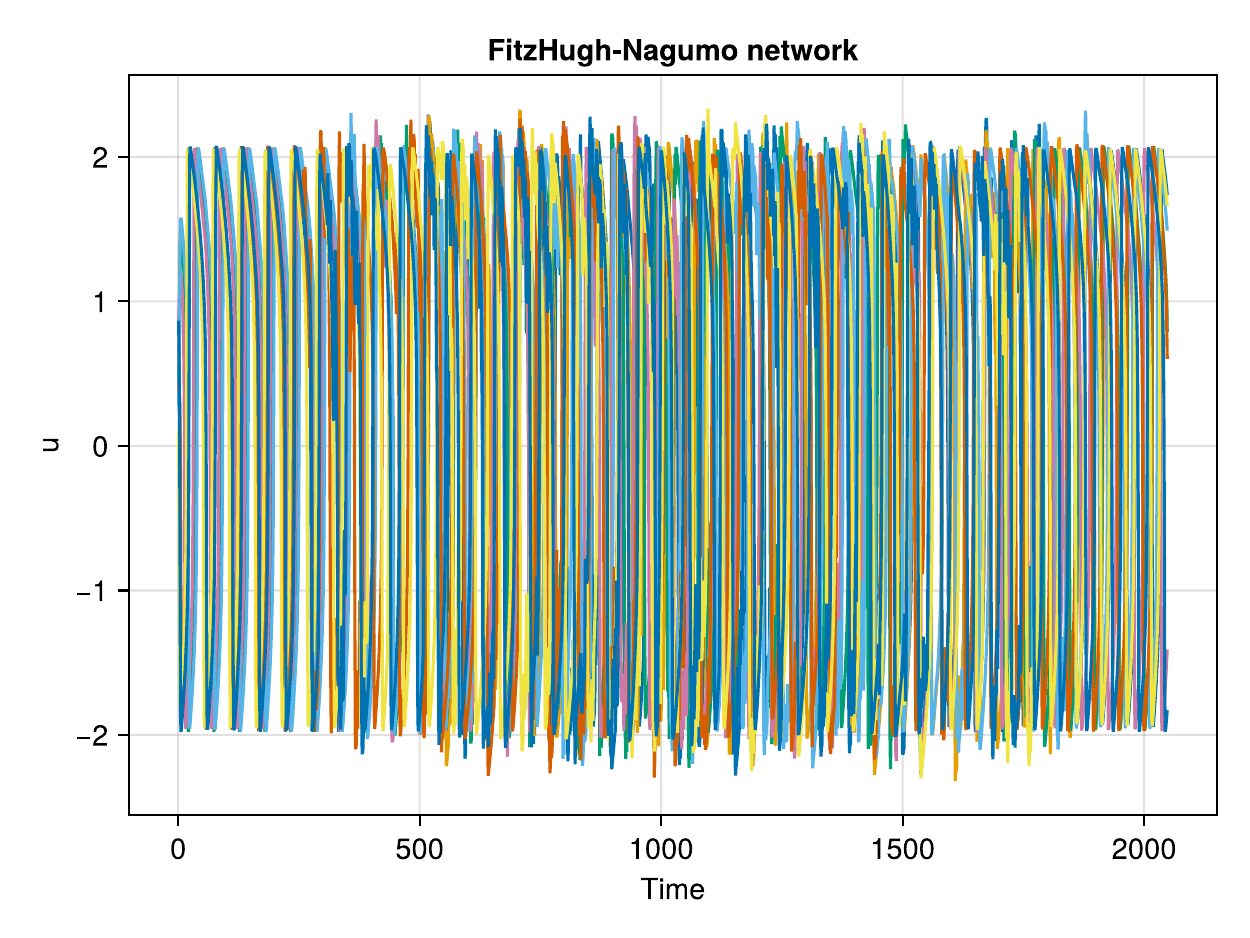}
    \caption{The time evolution of the network of FHN oscillators. The network topology corresponds to the one from Figure \ref{fig:fig1}, d). The following oscillators parameters are used to obtain this dynamics: FitzHugh-Nagumo $a=0.5, \sigma= 0.06, e_0 = 0.0, j_0= g(t)$. The network includes 1797 nodes and 3227412 edges, here we are displaying only 8 oscillators during time evolution of 2048 steps.}
    \label{fig:net_dynamics}
\end{figure}

\begin{figure}
    \centering
    \includegraphics[width=0.8\linewidth]{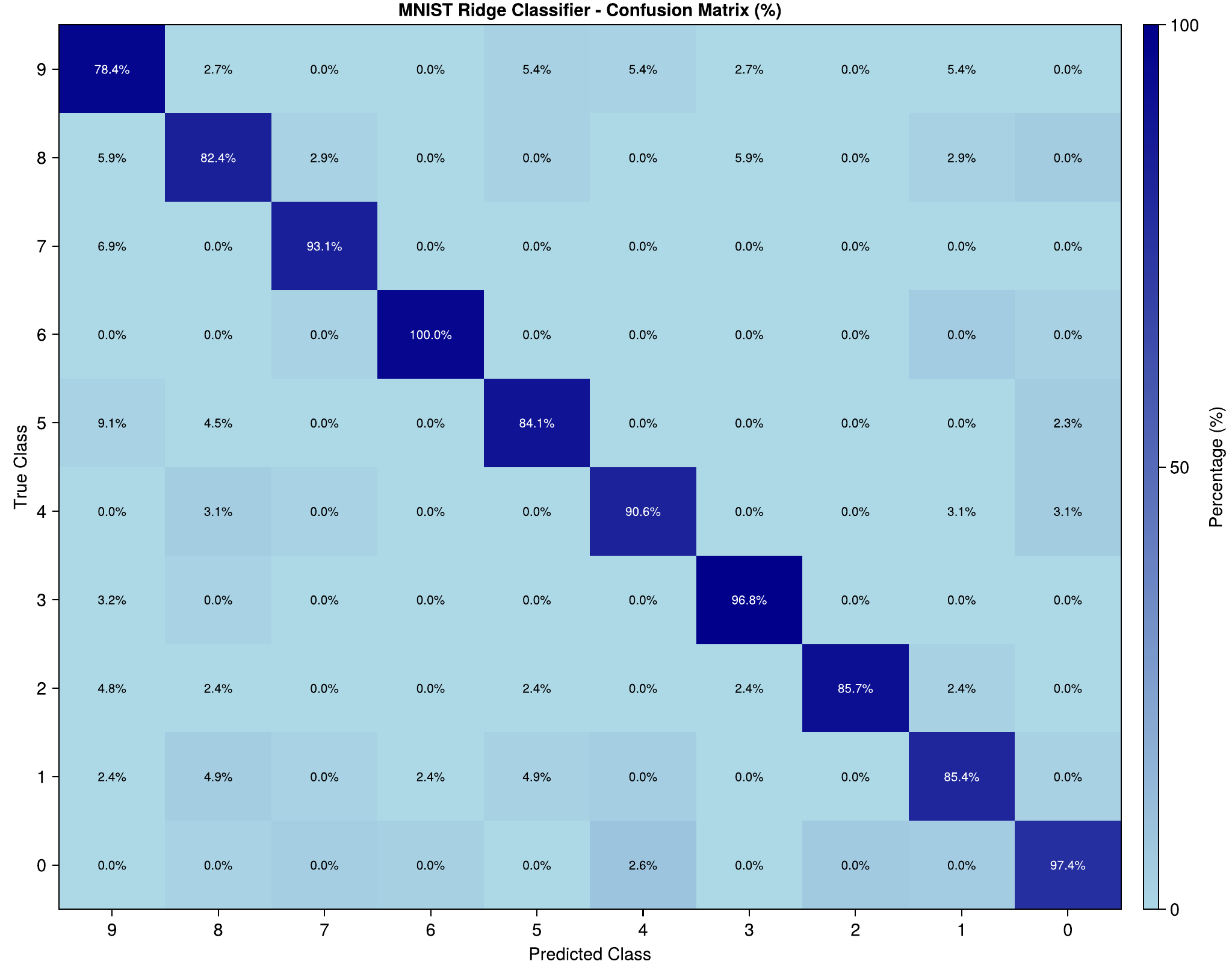}
    \caption{Confusion matrix for classification of scikit-learn dataset using Ridge Regression optimization.}
    \label{fig:mnist_cmat}
\end{figure}

\paragraph{Network dynamics}
Before discussing the experimental results, we aim to gain insight into the network's behavior from a dynamical perspective. To this end, we present the free motion of different oscillators networks with and without distortion (where distortion corresponds to data feeding as described in section \ref{sec:network_data_feed}). The network dynamics without distortion are illustrated in Figure \ref{fig:net_dynamics}. As seen for {\em Kuramoto} networks, the dynamics is characterized by chaotic behavior of each individual oscillator with no prevailing dynamic mode and without obvious structure. Interesting case can be seen for {\em FitzHugh-Nagumo} network, where an ordered dynamics with synchronized oscillations is seen. This ordered state of entire network, known as supercritical, is described in the work of Al Bettie at el. \cite{al_beattie_criticality_2024}. However, such behavior is strongly dependent on the choice of oscillators internal parameters, in particular the membrane potential in Eq. (\ref{eq:fhn}), (\ref{eq:rec_var}). Tuning of this parameter may drastically change the dynamics of the same network, bringing it to from the orderly behaviour, shown in Figure \ref{fig:net_dynamics}, to complete chaotic one, like Kuramoto cases in the same figure.
% figure network dynamics
The time evolution of the networks with different topologies (see Figure \ref{fig:fig2}, a-d) is similar to Figure \ref{fig:net_dynamics}. Due to the space limit, we do not present the dynamics for all cases, however the results can be reproduced using provided code repository.
The network dynamics of the nodes with input pulse train supplied can be seen in Figure \ref{fig:net_dynamics} for case of FHN-oscillators. The red vertical marker indicates the distortion, which enforces the starting condition of each individual oscillator in accordance with the input data values (see Section \ref{sec:network_data_feed}). For clarity, the graphs in Figure \ref{fig:net_dynamics} include the dynamics of several resonant and non-resonant oscillators. As can be seen, just shortly after distortion the resonating oscillators accumulate more energy that is characterized by greater osculation amplitudes (see red arrows on the graph). In long times periods the accumulated energy redistributes among the neighbors, bringing the dynamics of those resonant oscillators to the equilibrium state, analogues to one from Figure \ref{fig:net_dynamics}. Simultaneously, non-resonant oscillators do not distinctly react to the resonant. 
Additionally, it has to be noted that the dynamics of the network responds to the learned data patterns not simultaneously but reproduces the separate modes of the learned data during some time after distortion and before reaching the equilibrium state. This over time reconstruction of the patterns by the network may lead to some classification mistakes, which are discussed in the next section. 
\paragraph{Classification results}
The tests of the framework for case of reservoir readout were done for two datasets, namely i) scikit-learn digits patterns (see Figure \ref{fig:fig2}, e for examples), and ii) dry-bean dataset. Classification result for each dataset is discussed in subsections below.
%Data about scikit-learn digits set
\paragraph{Scikit-learn dataset}
Scikit-learn database used here was a down-sized version of the MINST dataset that contains 1797 samples of training image patterns of digits with 10 classes. Each image pattern was the size of 8x8 pixels, while the examples of those are shown in Figure \ref{fig:fig2}, e). The training of the network was organized as described in section \ref{sec:Network_training}, while the size of the training set included 1437 samples. The test sets included 360 samples that were not seen inside the training set. The training included the iteration over 1797 batches, while the test results can be seen in Figure \ref{fig:mnist_cmat}. As can be seen from the Figure \ref{fig:mnist_cmat}, the best classification accuracy reached was of $88\%$ for the system of FitzHugh-Nagumo oscillators. The reasons for inaccuracies can be transformation of the pixeled image to spike-train data. The overall analysis of all misclassifications showed that most frequent mistakes were obtained by classifying digits with similar patterns as visible from
confusion matrix see Figure \ref{fig:mnist_cmat}. The most vivid example is digits 1 and 4 (see Figure \ref{fig:fig2}e) for an example).

Additionally, to results in Figure \ref{fig:mnist_cmat}, we would like to mention the nature of misclassifications errors inside the network. As mentioned above, the network reproduces the different modes of the learned patterns in time (but not all at once). 
%A representative example can be seen in Figure 6, where the network with FitzHugh-Nagumo oscillators reproduces in time the separate modes (Figure 6, b-c) of the original image (Figure 6, a). The modes are excluding the patterns the intensity of which is not high. In Figure 6, c the marker highlights the pixel that was not present in any of the modes. This corresponds to learning of only stronger patterns, losing non-significant details and therefore bringing to classification errors. For Kuramoto cases the reproduction of the modes takes place in similar way.
% figure 4 confusion matrix written in precent 
% figure 6
\paragraph{Dry-bean dataset learning}
The second dataset used is drybean dataset with 16 attributes. The 16 attributes of the drybean dataset can be directly fed to the 16 nonlinear oscillators of our network.

Our reservoir is made up of 64 FitzHugh-Nagumo (FHN) oscillators. We converted the dry bean dataset, which contains 16 features, into spike trains and input these into the network of oscillators. We employed a complete graph configuration for reservoir computing and experimented with the Watts-Strogatz graph topology for classification tasks. The edge weights between the nodes were adjusted to operate close to the chaotic regime, as suggested by previous studies. Supervised learning was conducted over 5,000 epochs, using categorical cross-entropy as the loss function and optimizing it with the BFGS method. The classification accuracy on the test samples reached $92.3\%$. This performance is comparable to other methods reported in the literature and in previous studies involving FHN networks.
\begin{figure}
    \centering
    \includegraphics[width=0.8\linewidth]{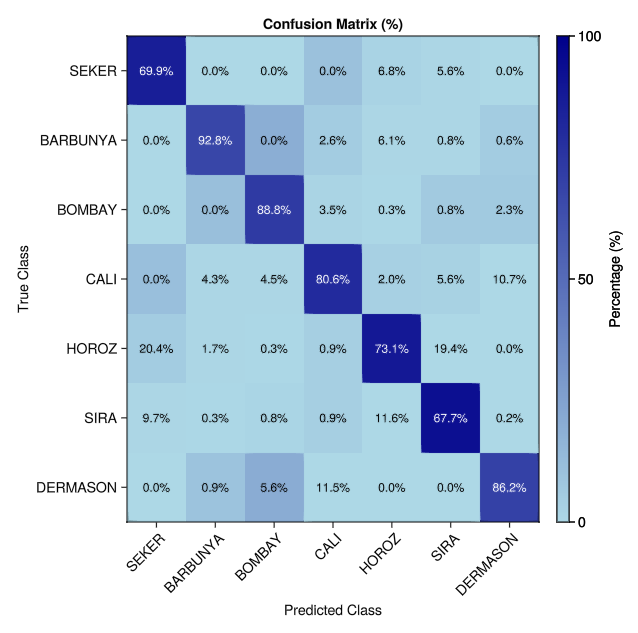}
    \caption{Confusion matrix for classification of the dry bean dataset}
    \label{fig:mnist_cmat_dry_bean}
\end{figure}
In our study, we utilized a fully nonlinear network of FHN oscillators with various graph topologies. While a systematic comparison across topologies is not undertaken, empirical results suggest that reservoirs with higher inter-domain connectivity (such as all-to-all and Watts-Strogatz) outperform more locally clustered or modular structures such as Barabási-Albert graphs. This finding is consistent with the hypothesis that enhanced coupling facilities richer mixing of spatiotemporal features and improves separability at the readout.

\begin{figure}
\centering
    \includegraphics[width=1\linewidth]{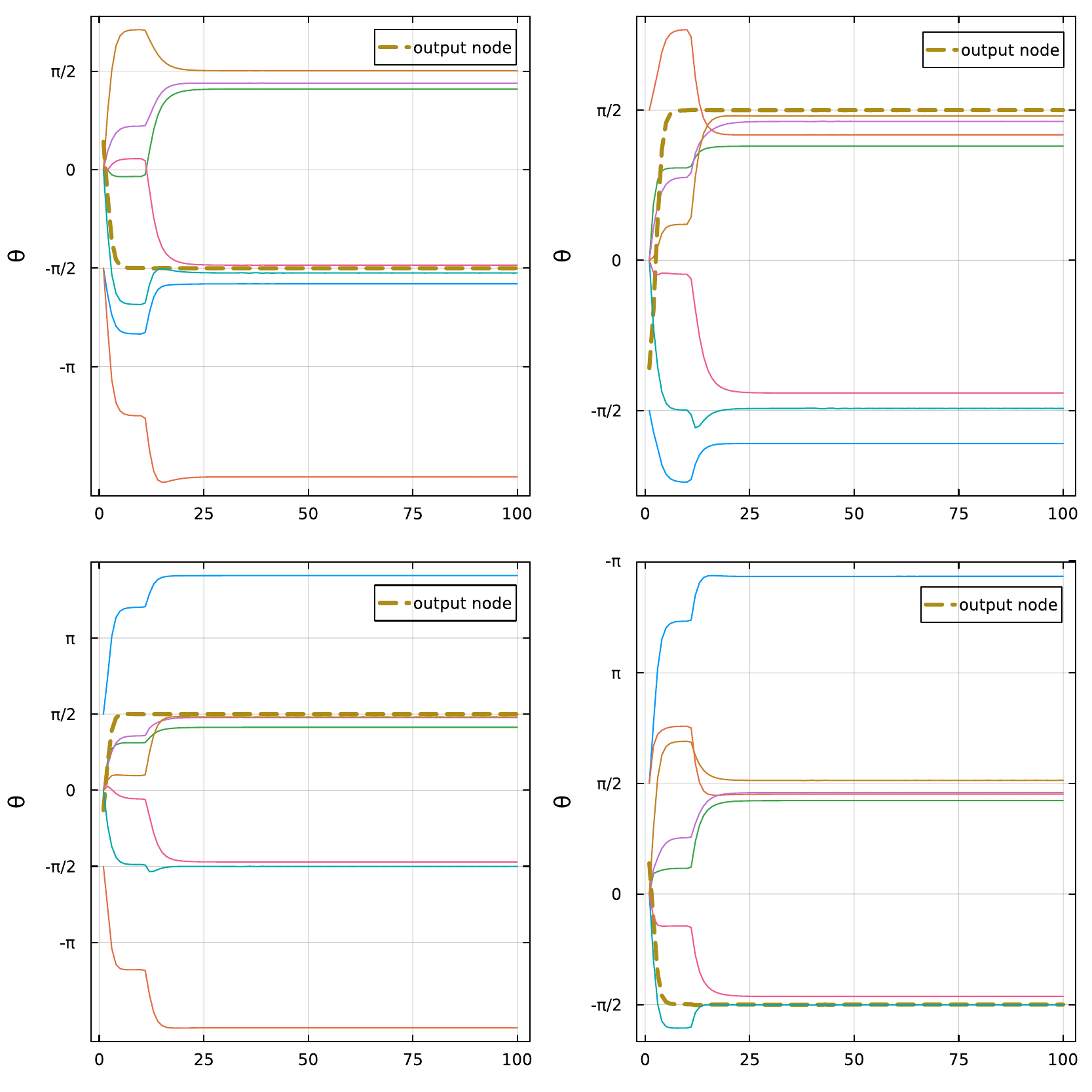}
    \caption{The dynamics of the network of oscillators in Watts-Strogatz topology showing learning of XOR gate in 8 node network. Note that output node 5 correctly predicts ($-\pi/2, \pi/2, \pi/2, -\pi/2$).}
    \label{fig:xor_erdos_renyi}
\end{figure}

\begin{figure}
    \centering
    \includegraphics[width=1 \linewidth]{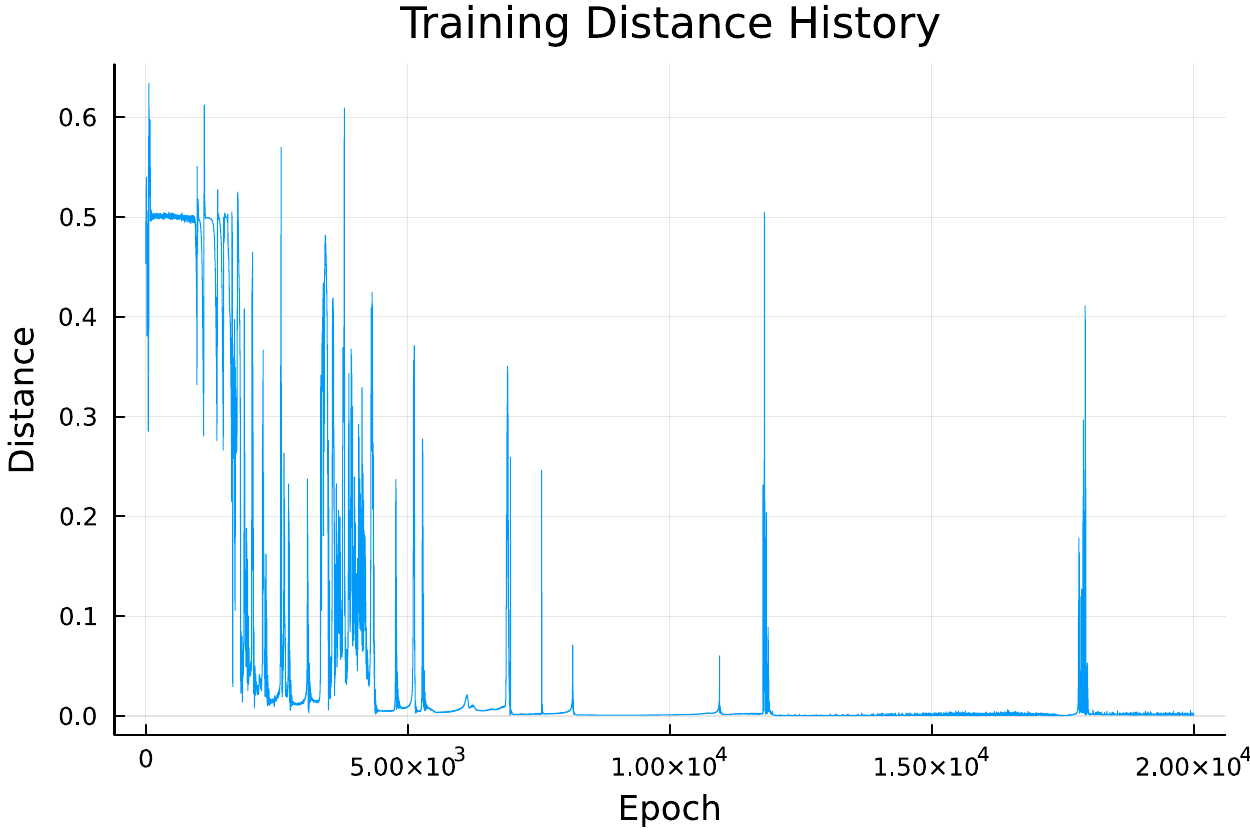}
    \caption{Distance function for 5-oscillator Kuramoto network for training over epochs.}
    \label{fig:five_osc}
\end{figure}

\paragraph{Learning the XOR gate}
The XOR gate is a task that requires at least one hidden layer \cite{goodfellow_deep_2016}, making it a pivotal example in the history of machine learning. In this section, we follow the procedure outlined by Wang \emph{et al.} \cite{wang_training_2024} The true and false values of the XOR gate are represented as follows for Kuramoto - $\phi = \pi/2$ and $\phi= -\pi/2$ (see Eq.(\ref{eq:kuramoto})). 
We train 5-oscillator a fully connected network for a duration of time interval $T$. The final time is determined by observing the equilibrium point where the system converges to fixed points. The network's weights are initialized from a normal random distribution, and oscillator's initial conditions are also drawn from a normal random distribution. The universal differential equations \cite{rackauckas_universal_2021} are used to represent the network activity and are defined in accordance with Eq.(\ref{eq:kuramoto}) and Eq.(\ref{eq:coup_oscil}):
\begin{equation} 
   \label{eq:kuramoto_nn}
   \frac{d\phi_i}{dt}=\omega_i+NN(\phi; \theta) + h \sin(\phi_i-\psi_i) 
\end{equation}
Where $NN(\phi;\theta)$ is shallow neural network with parameters $\theta$. The neural network is all-to-all, Watts-Strogatz, and Erd\H{o}s-R\'enyi connected network that learns interaction between nonlinear oscillators in the system to produce desired outcome at output nodes. 
For optimization, we use the Sophia solver \cite{liu_sophia_2024} for the first 100 epochs, then switch to the BFGS optimizer until convergence. Following Wang \emph{et.al.} \cite{wang_training_2024}, we also record the distance function. For Kuramoto oscillators this function is defined as:
\begin{equation}
    \label{eq:distance_func}
    D(\phi_i, \phi_i^t) = \sum_i 1 - \cos(\phi_i-\phi_i^t)
\end{equation}
where, $\phi_i^t$ represents the target phase, while $\phi_i$ is the current phase of the $i$-th oscillators (Eq.\ref{eq:kuramoto}). Although the distance function described in Eq.(\ref{eq:distance_func}) can be used for training, in practice as shown by Wang \emph{et. al.} \cite{wang_training_2024} the cost function $C(\phi_i,\phi_i^\tau)=\sum_i[-\ln(1+\cos(\phi_i-\phi_i^\tau ))]$ works better as it avoids the unstable solutions which differ by $\pi$ from the true solution.The dynamics of the oscillators for this task are illustrated in Figure \ref{fig:xor_erdos_renyi}, where the case for Kuramoto model is presented. 
Using Equilibrium Propagation (EP), we train 5-oscillator Kuramoto network with all-to-all connectivity to perform XOR task. For random initial conditions, Fig.\ref{fig:five_osc} shows the distance function which converges to zero. Proving that Kuramoto network learns XOR function using EP training.
%%figure for EP training using 5-osc Kuramoto network
This application suggests possibility of the hybrid systems, where part of the network can be trained using the EP method on neuromorphic platform and smaller part of the network that uses universal differential equation could be trained on the standard silicon-based hardware.

\subsection{Other applications}
As an outline of this work, we would like to present two other applications that are extensions for our framework that require its slight modification. Those are i) patterns recognition and ii) dynamic system identification.
\subsection{Pattern recognition using Hebbian learning}
\begin{figure}
    \centering
    \includegraphics[width=0.8\linewidth]{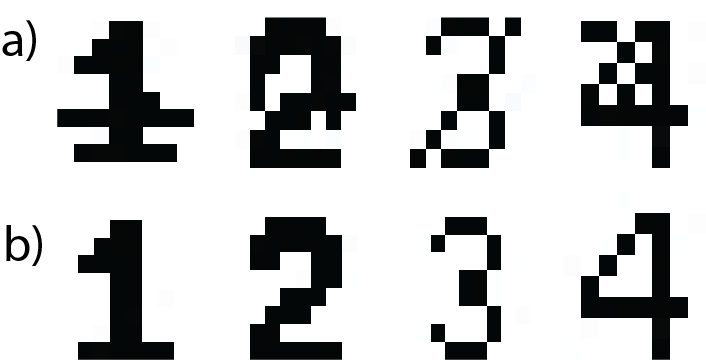}
    \caption{a) Test patterns to recognize hand written numbers b) Patterns recognized with fully connected network of 8x8 Kuramoto oscillators recognize the patterns after forcing period $T_f > 400$.}
    \label{fig:rec}
\end{figure}

\begin{figure}
     \centering
     \includegraphics[width=1\linewidth]{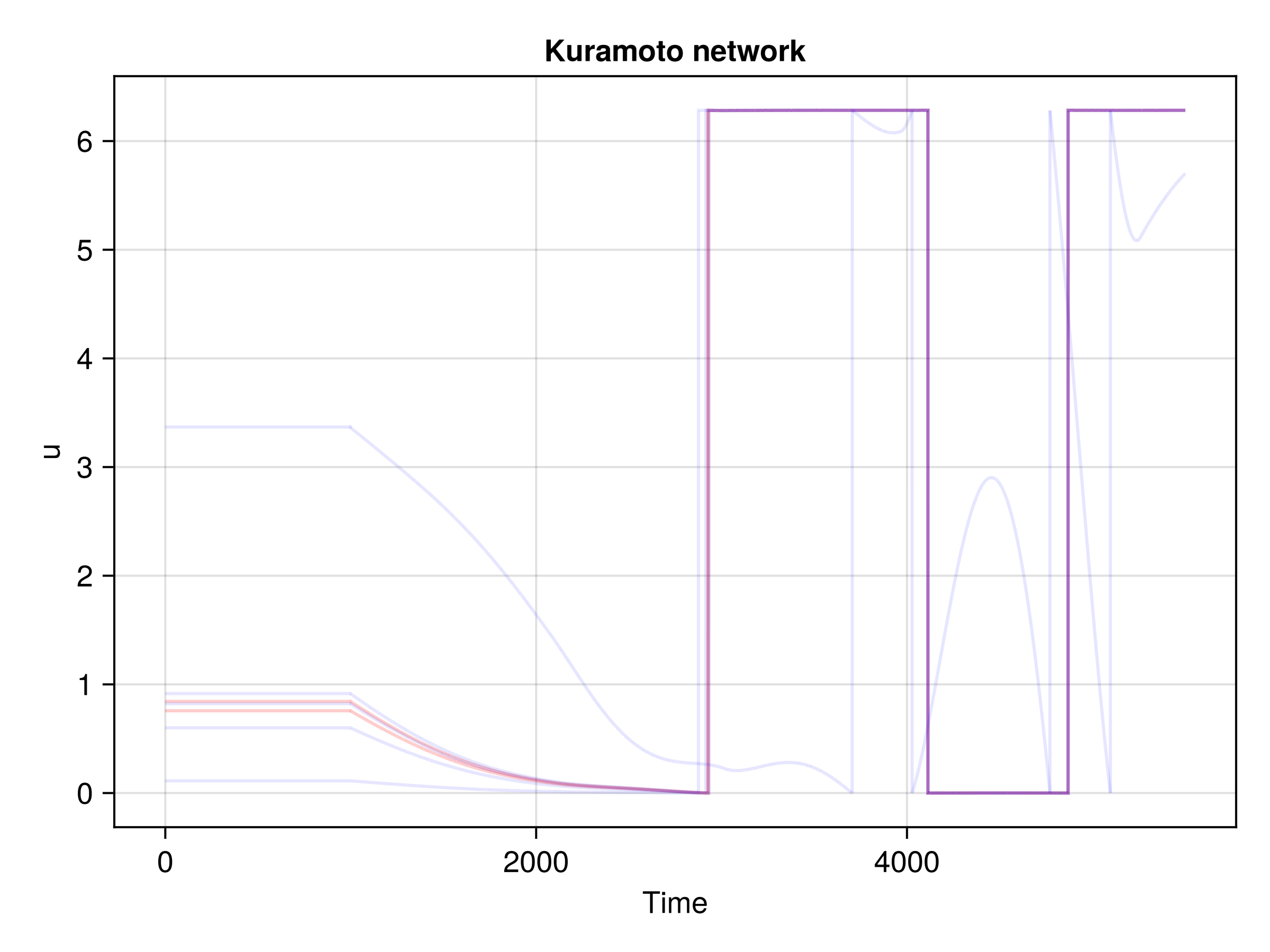}
     \caption{The time evolution during pattern recognition by the network of Kuramoto oscillators using Hebbian learning rule Eq.(\ref{eq:hebian}) applied to number pattern 1. Note: parameters used $\omega=1.0, \epsilon=100, T_f=400, h=0.1$}.
     \label{fig:4}
 \end{figure}

In this section we will explore several applications of nonlinear oscillator network in machine learning tasks. Pattern recognition and associative recall has a long history in neuro-computing research \cite{hopfield_neural_1982, izhikevich_phase_2000}. Pattern recognition using the network of the lasers was discussed by Hoppensteadt and Izhikevich \cite{hoppensteadt_synchronization_2000, hoppensteadt_synaptic_1996}. The oscillatory neurons are connected via phase locking between different units. In this model the lasers interact via phases, and data are encoded using phase modulation. The phase model follows the standard Kuramoto framework, as discussed in previous section. This model is equivalent to a {\em Hopfield} network with symmetric connections $k_{ij}=k_{ji}$ and antisymmetric $\psi_{ij}=-\psi_{ji}$, which has well defined potential energy:
%%potential energy definition
\begin{equation}
    \label{eq:ene_def}
    \phi_i^{'} = \frac{\partial E}{\partial \phi_i}
\end{equation}
where,
\begin{equation}
    \label{eq:ener_kuramoto}
    E(\phi_i) = -\frac{1}{2}\sum_{ij} k_{ij}cos(\phi_i +\psi_{ij} - \phi_j)
\end{equation}
During the evaluation of the network, the input of the selected units is fixed by components of the input pattern, which in our example represents pixel values of an image. A similar objective can also be achieved by external driving of the network. The system eventually stabilizes at its equilibrium, corresponding the lowest energy state.
The output values of the oscillators in the network represents the phase information of the output vector. The Hebbian learning rule is employed to memorize the $m$ complex input vectors:
\begin{equation}
 \label{eq:complex_v}
\xi^k = (\xi_1^k,\xi_2^k,...,\xi_n^k) \in C^n, \quad \|\xi^k\| = 1
\end{equation}
where $k=1,\dots,m$. 
The $\xi_i^k$ and $\xi_j^k$ are complex vectors and angle differences between $i$ and $j$-th nodes in the network are given by $\phi_i - \phi_j$. If two oscillators are in phase, we have $\xi_i^k=\xi_j^k$ and when they are in antiphase, $\xi_i^k=-\xi_j^k$. Following Hoppensteadt-Izhikievich's approach, we apply the complex {\em Hebbian} learning rule \cite{hoppensteadt_synaptic_1996-1}:
\begin{equation}
    c_{ij} = \frac{\epsilon}{n}\sum_{k}^{m}\xi_i^k\bar{\xi_j^k}
    \label{eq:hebian}
\end{equation}
with,
\begin{equation}
     c_{ij} = k_{ij} e^{\psi_{ij}} 
\end{equation}
Here, $\epsilon>0$ is the strength of the connections between the oscillators. The network initialization in this example is done by applying a strong periodic signal to the connection free oscillators, which forces them to lock to the input signal. After the initialization stage, the connections between oscillators are restored, allowing network to converge to a cycle attractor.
For the memorizing task, we used a the small $8x8$ dataset of digits encoded as binary image $\psi=\{\textpm{1},\textpm{0}\}$. This dataset is available through the ScikitLearn package \cite{pedregosa_scikit-learn_2011}. The sample consists of $8x8$ greyscale images of digits from $0$ to $9$. A sample figure of this dataset was provided in the previous section. To execute this task, we used standard all to all graph of same size as input data. Figure \ref{fig:rec} shows the results of the recognition task using the Hebbian learning rule. 
 This example demonstrate that a weakly coupled network of Kuramoto oscillators can recognize complex patterns (see Figure \ref{fig:4}) and form oscillatory memory traces. However, it remains unclear whether comparable performance can be achieved with locally connected networks. As argued by Wang \emph{et. al.} \cite{wang_training_2024}, fully connected architectures appear necessary for digits classification tasks due to difficulties in integrating information across different spatially separated input regions. To probe this issue, we evaluated multiple network topologies, including Erd\H{o}s-R\'enyi and Watts-Strogatz graphs. In our experimental results the additional networks of type suggested in Figure \ref{fig:fig1} were used. Both Erd\H{o}s-R\'enyi and Watts-Strogatz topologies can be used for recognition task but the connectivity of graph should sufficient to store the information for recovering the memorized data from $c_{ij}$ connections. According to Hoppensteadt and Izhikevich \cite{hoppensteadt_synchronization_2000} when $c_{ij}$ are synaptic coefficient for a {\em Hopefield-Grosseberg} neural network, then the network contains $2m$ attractors in some small neighborhoods of $\xi^k$ and their mirror images $-\xi^k$. The maximum number of images is given by $m < n/8$ under assumption that $\xi^k$ are nearly orthogonal to each other.

 %% dynamic system identification
 \subsection{Dynamic system identification}
\begin{figure}
    \centering
    \includegraphics[width=0.8\linewidth]{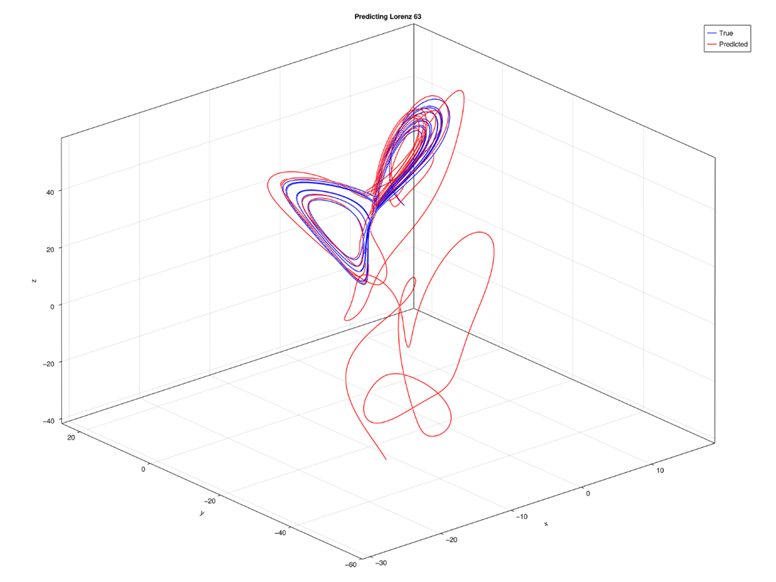}
    \includegraphics[width=1\linewidth]{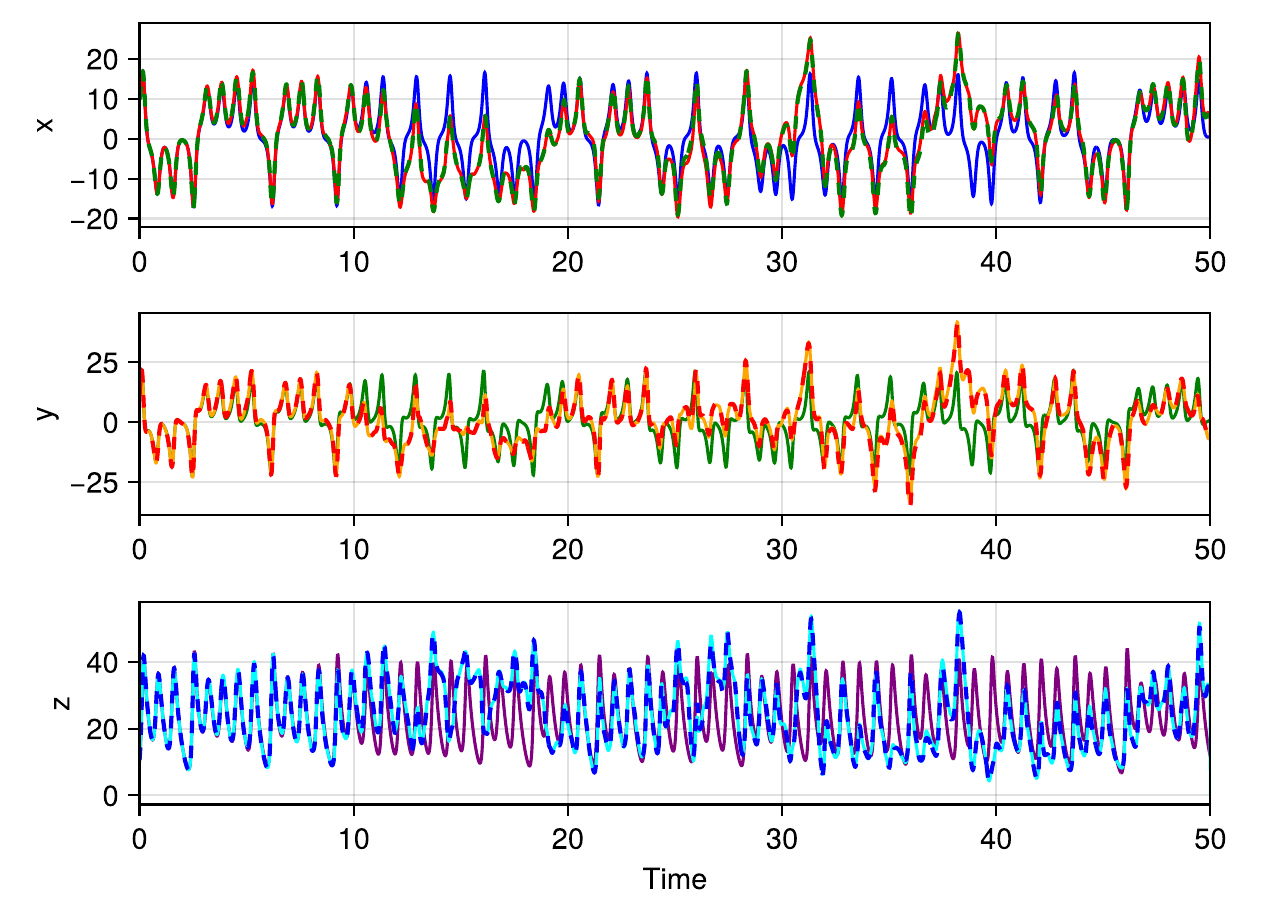}
    \caption{a) Lorenz system with FHN-oscillators network with activation function 
    $swish(x) = x * sigmoid(x)$ with the size of the network with dimension $d = 64$ and density of $0.25$ 
    b) The performance for the attractor is very good.}
    \label{fig:lorenz_attr}
\end{figure}

We are simulating the \emph{Lorenz63} system 48 in RC configuration using the complete graph. The Lorenz system consist of three coupled ordinary differential equations, with parameters $\sigma=10, \rho=28, \beta=8/3$:
\begin{align}
    \dot{x} &= \sigma(y-x) \\
    \dot{y} &=  x(\rho-z)-y \\
    \dot{z} &=  xy -\beta z
\end{align}
The Lorenz equations are benchmark problem for demonstrating chaotic behavior \cite{lorenz_deterministic_1963}. The chaotic attractor in the phase space ${x(t),y(t),z(t)}$ forms the famous butterfly like structure, as shown in Figure \ref{fig:lorenz_attr}.
The reservoir computation uses a complete graph with $N =64$ FHN-oscillators, providing a total of 128 degree of freedom (DOF). We only use output voltages $u_k$ as computational DOFs and output layers matrix $W_{out}$ is trained using the 2-layer network (64,256)(256,3). The two layer neural network is optimized using the BFGS algorithm with standard parameter values.
% figure
Due to chaotic nature of the Lorenz system, we observe good approximation in the short-term, but the error grows rapidly as the initial inaccuracies accumulate. Our implementation performs on par with standard nonlinear (RC) using a $\tanh(x)$ activation function and quadratic readout matrix $[r, r\circ r]$, as described by Bollt \cite{bollt_explaining_2021}.
\section{Conclusions} %%checked
Our study explores a hybrid architecture that combines networks of chaotic oscillators with machine learning for pattern recognition in signal streams. In the proposed framework, machine learning approximates the oscillators' coupling terms, thereby preprogramming the network expected responses to anticipated patterns. Experimental results demonstrate that this ML approximation approach is versatile, applied to a wide range of network topologies and oscillator types, and enables the rapid design of chaotic networks through straightforward training procedures. 

Using chaotic oscillators as the foundational computational units of these networks holds significant promise for harnessing the full potential of chaos in signal processing and dynamical systems prediction. In particular, the ESNs and RC framework enable memory retention and delay embedding, which are particularly beneficial for forecasting dynamical systems. Reservoir computing is more amenable to training because only the output layer requires optimization using backpropagation or direct ridge regression method. Moreover, linear reservoir computing with nonlinear output layer permits a least squares solution that is computationally far more efficient than the backpropagation employed in traditional neural network training. Our implementation using the FitzHugh-Nagumo (FHN) network yields a better mean square error, primarily attributable to its more accurate reproduction of the phase trajectory over longer time intervals. This enhancement improves pattern recognition, even under high noise-to-signal ratios and complex data encoding scenarios.

It is important to note that this work does not fully realize the potential of nonlinear oscillators as a general-purpose neuromorphic system. However, in conjunction with other related research, it serves as a preliminary step toward developing a new generation of machine learning systems. Future research could build on this foundation by incorporating a broader range of machine learning modalities in chaotic oscillator networks, such as supervised, unsupervised, and reinforcement learning, as well as exploring more complex architectures. This paper is accompanied by a code repository that provides access to many additional results not included in the manuscript: [github\url{}].

%\bibliographystyle{plain}  
%\bibliography{NonlinearOscillators,NonlinearOscPaper}

\printbibliography
\end{document}